\documentclass[showpacs,floatfix,amsmath,superscriptaddress,amssymb,twocolumn,prb]{revtex4}
\usepackage{latexsym}
\usepackage{graphicx}
\begin{document}
\title{Quasiparticle scattering and local density of
states in the $d$-density wave phase}
\author{Cristina Bena}
\affiliation{Department of Physics, University of California
at Santa Barbara, Santa Barbara, CA 93106, USA}
\author{Sudip Chakravarty}
\affiliation{Department of Physics and Astronomy,
University of California Los Angeles, Los Angeles, California 90095-1547, USA}
\author{Jiangping Hu}
\affiliation{Department of Physics and Astronomy,
 University of California Los Angeles, Los Angeles, California 90095-1547, USA}
\author{Chetan Nayak}
\affiliation{Department of Physics and Astronomy,
University of California Los Angeles, Los Angeles, California 90095-1547, USA}
\date{\today}

\begin{abstract}
We study the effects of single-impurity scattering on the local
density of states in  the high-$T_c$ cuprates.   We compare the
quasiparticle interference patterns in three different ordered
states: $d$-wave superconductor (DSC),  $d$-density wave (DDW), and
coexisting DSC and DDW (DSC-DDW). In the coexisting  state,
at energies below
 the DSC gap, the patterns are almost identical to
 those in the pure DSC state with the same DSC gap.
However, they are significantly different for energies greater than or equal to
the DSC gap. This transition at an energy
around  the DSC gap can be used to test the nature of the
superconducting state of the underdoped cuprates by scanning tunneling
microscopy.  Furthermore, we note that in the DDW state the effect
of the coherence factors is stronger than in the DSC state. The
new features arising due to DDW ordering are discussed.
\end{abstract}
\maketitle

\section{Introduction}

One of the intriguing features of the cuprate high temperature
superconductors is the existence of a pseudogap phase
in the
normal state.\cite{Timusk99}
Elucidating the nature of the pseudogap would be an
important step in understanding the physics of high-temperature
superconductivity. Among many scenarios,
one concrete proposal is that the pseudogap
is due to a hidden broken symmetry \cite{Chakravarty01} of $d_{x^2-y^2}$ type in the particle-hole
channel.\cite{ddw} In this paper we examine the consequences of the
presence of DDW ordering on the quasiparticle interference patterns
observed by scanning tunneling measurements (STM).

Recently, interesting STM
measurements\cite{Hoffman1,Hoffman2,Howald} on Bi-2122
have been performed to obtain Fourier transformed spectra
of the local density of states. This technique is
called Fourier transform
scanning tunneling spectroscopy (FT-STS). So far, the experiments
have been carried out  in the superconducting state at low temperatures, and
the results can be interpreted as interference patterns
due to elastic scattering of the quasiparticles from impurities.
\cite{Byers,Flatte,Hoffman1,Wang03} Additional support of this interpretation comes
from a more recent work in Ref.~\onlinecite{Scalapino}.
The scattering of quasiparticles between regions of the Brillouin zone
with high densities of states yields peaks in FT-STS. In a $d$-wave
superconductor (DSC), the regions of high local density of states
(LDOS) are situated
at the tips of the banana-shaped contours
of the quasiparticle excitation spectrum. \cite{Hoffman1} The
wavevectors  of the observed peaks are consistent
with the wavevectors that connect the tips of these bananas
(see Fig.{\ref{surface}).
Thus, the experimental results are consistent with the known
pictures of the Fermi surface  and the LDOS in
the pure DSC state.
Here, we investigate the possible form of STM spectra in
both the superconducting
and normal states from the perspective of DDW.
The T-matrix formalism, \cite{Wang03}
combined with a numerical analysis, are used to obtain
the quasiparticle interference patterns of the various ordered states.

We first consider a pure DSC state and recover results similar to Ref.
\onlinecite{Wang03}. Though  the  band structure we used is different from theirs,
the main features such as the emergence of peaks at particular wave vectors
are still observed. However, as we will discuss later, differences in
the band structure can have large effects on the details of the STM spectra, especially due
to coherence factors.  Fortunately, none of our robust qualitative conclusions depend on the spurious sensitivity to the band structure.

Our main focus is to study the mixed, or coexisting, DSC-DDW and the pure DDW
states. According to the DDW theory,\cite{Chakravarty01} the mixed DSC-DDW and the pure DDW characterize the underdoped superconducting and the pseudogap phase
respectively. We first analyze the mixed state and find that if we
probe the system at energies larger than the DSC gap, significant
differences among the interference patterns emerge compared to a
pure DSC state, while at energies below the DSC gap, the patterns
are almost identical. For example, let us focus on the position of
the peaks along the $(\pi,\pi)$ direction in the interference
patterns. For a pure DSC state they should move away from the
origin with increasing energy and eventually get close to the
$(\pi,\pi)$ point for energies of the order of the gap. If the
 state is a mixed state (DSC-DDW) with two
different order parameters, the total gap is roughly equal to the square
root of the sum of the two gaps square. In this case the same
peaks should shift away from the origin faster (for lower energy)
and get close to the
 $(\pi,\pi)$ point at energies  of the order of the DSC gap
  which is typically much smaller
than the total gap. For energies larger than the DSC gap, the main
features of the STM spectra is  significantly different from those
in the  pure DSC state. Therefore, the energy of the DSC gap in
the mixed states marks a transition in the interference patterns.
The effect becomes less pronounced
 with an increase in doping, as the DSC gap is getting closer to the
total gap of the system.
Based on this
observation we believe that the FT-STS obtained below the
superconducting transition temperature in the underdoped cuprates
may reveal important information about the nature of the order parameter.

In the pure DDW state, the interference patterns reveal features
similar to the ones observed in the DSC state and others that are
entirely new. These features could be used to detect DDW
order. One of the important differences from the
case of DSC order is that the DOS is not as strongly
peaked in the DDW state, so the STM spectrum is
more strongly dependent on coherence factors.  These depend
not only on the order parameter and on the band structure, but also
on the details of the scattering and on the type of impurities.
We illustrate this effect by comparing the interference patterns
for two parameter sets which are identical save for the sizes of
their DDW gaps. Changing the size of the DDW gap has relatively
little effect on the equal energy contour plots.
However, the interference patterns are dramatically different due to the
difference in the coherence factors.
The authors of Ref. \onlinecite{Pereg-Barnea03}
emphasized their importance. We show that
their observation is pertinent and, in fact, explains why
their lowest-order Born approximation results miss some of the physics
captured in the full T-matrix approach. Also, the strong dependence
of the STM spectra on coherence factors suggests that they
could be used to extract detailed information about
the DDW gap, band structure, etc.,  which could be compared
with angle-resolved photoemission spectra (ARPES) data.

The paper is organized as follows. In section II we describe
our model and the analytic formalism. In section III, we describe
the numerical analysis and discuss the results; we conclude in
section IV.

\section{Analytic formalism}
In the presence of DDW and DSC order, the general  mean field
Hamiltonian for a high $T_c$ superconductor is given by
\begin{widetext}
\begin{eqnarray}
H=\sum_{k\sigma}[\epsilon(k)-\mu]c_{k\sigma}^+c_{k\sigma}+\sum_{k}
\Big[\sum_{\sigma}
iW(k) c_{k\sigma}^+c_{k+Q,\sigma} + \Delta(k)
c_{k\uparrow}^+c^+_{-k\downarrow}\Big] +h.c.
\end{eqnarray}
\end{widetext}
where $W(k) = W_0(\cos k_x- \cos k_y)/2$ and
$\Delta(k) = \Delta_0(\cos k_x-\cos k_y)/2$ are the DDW order
(see also ref. \onlinecite{ddw}) and DSC order
parameters respectively, and $Q=(\pi,\pi)$.
The sums over $k$ include all the wave vectors in the first Brillouin zone
(BZ), $|k_x| \le \pi$, $|k_y| \le \pi$. We take the lattice constant $a=1$.

The kinetic energy is $\epsilon(k)
=\epsilon_1(k)+\epsilon_2(k)$, where $\epsilon_1/\epsilon_2$ are respectively
  odd/even with respect to the shift of the wavevector by $Q$;
$\epsilon_1(k)=-\epsilon_1(k+Q)$ and
$\epsilon_2(k)=\epsilon_2(k+Q)$.   In the numerical calculation,
we focus on a simple $t,t'$  band structure  with dispersion
$\epsilon_1(k)=t (\cos k_x+\cos k_y)/2$ and $\epsilon_2(k)= t'
\cos k_x \cos k_y$.

To simplify the calculation, we
introduce a four-component (spinor) field operator, $\psi_k^+
=(c_{k\uparrow}^+,c_{k+Q\uparrow}^+,
c_{-k\downarrow},c_{-k-Q\downarrow})$ \cite{Nambu}. In this new basis,
 the above Hamiltonian can be written as
\begin{eqnarray}
H=\sum_k\psi_k^+A(k)\psi_k,
\end{eqnarray}
where $k$ is summed over half of the original
Brillouin zone (reduced Brillouin zone - RBZ),
namely, $|k_x|+|k_y|\leq \pi$. $A_k$ is a four by
four matrix given by
\begin{widetext}
\begin{eqnarray}
A_k= \left( \begin{array}{cccc}
 \epsilon_1(k)+\epsilon_2(k)-\mu & iW(k) & \Delta(k) & 0 \\
-iW(k) &-\epsilon_1(k)+\epsilon_2(k)-\mu &  0 & -\Delta(k)  \\
 \Delta^*(k) & 0 &-\epsilon_1(k)-\epsilon_2(k)+\mu & iW(k)  \\
    0 &-\Delta(k)^* &-iW(k) &\epsilon_1(k)-\epsilon_2(k)+\mu
    \end{array} \right).
\end{eqnarray}
\end{widetext}
The eigenvalues of $A_k$ are $\pm E_1(k)$ and $\pm
E_2(k)$, where
\begin{eqnarray}
E_{1,2}(k)& = &\{(\sqrt{\epsilon_1(k)^2+W(k)^2}\pm
[\epsilon_2(k)-\mu])^2+\Delta(k)^2\}^{\frac{1}{2}} \nonumber \\&&
\end{eqnarray}
We consider impurity scattering of the form:
\begin{eqnarray}
H_{\rm imp}= \sum_{k,k'\,}
\sum_{\, \alpha,\beta =\uparrow,\downarrow}
V_{kk'\alpha\beta}c^+_{k\alpha}c_{k'\beta}
\end{eqnarray}
Up to a constant,  we can write it as
\begin{eqnarray}
H_{\rm imp}=\sum_{k,k' \in RBZ}\psi_{k}^+ V(k,k')\psi_{k'},
\end{eqnarray}
where $V(k,k')$ is a four by four matrix.

One can define a finite temperature (imaginary time) Green's function,
\begin{eqnarray}
G(k_1,k_2,\tau)=-{\rm Tr}\left[ e^{-\beta(K-\Omega)}\, {\rm T}_{\tau} \,
\psi_{k_1}(\tau)\psi^+_{k_2}(0)\right]
\end{eqnarray}
where $K=H-\mu N$, $e^{-\beta\Omega}={ \rm Tr}\, e^{-\beta K}$, and
$ {\rm T}_{\tau} $ is the imaginary time ordering operator. The
impurity scattering problem can be solved by computing the Fourier
transform of the Green's function from the $T$-matrix formulation
:
\begin{eqnarray}
G(k_1,k_2,i\omega_n)
=G_0(k_1,i\omega_n)T(k_1,k_2,i\omega_n)G_0(k_2,i\omega_n),
\end{eqnarray}
where
\begin{eqnarray}
G_0(k,i\omega_n)^{-1}= i\omega_n I-A_k,
\end{eqnarray}
$I$ is the $4 \times 4$ identity matrix, and
\begin{widetext}
\begin{eqnarray}
T(k_1,k_2,i\omega_n) = V(k_1,k_2)+\sum_{k'\in RBZ}V(k_1,k')
G_0(k',i\omega_n)T(k',k_2,i\omega_n).
\label{t}
\end{eqnarray}
\end{widetext}
For simplicity, we take the impurity scattering potential to be a
delta function
so that the matrix $V$ is independent of $k$
and $k'$; $V(k,k')=V$. For this case, we can solve Eq. (\ref{t})
and obtain
\begin{eqnarray}
T(i\omega_n)= [1-V\int dk \, G_0(k,i\omega_n)]^{-1}V
\end{eqnarray}
where the only difference from the standard result is that the
integral over $k$ is over the RBZ.

Consequently, the  local density of states $\rho(q,\omega)$, is given by
\begin{eqnarray}
\rho(q,\omega)&\thicksim&\frac{i}{2\pi}\sum_{k\in
RBZ}g(k,q,\omega). \label{g1}
\end{eqnarray}
where $g(k,q,\omega)$
 is defined as follows. Let $k'=k+q$.
If $k'$ is in the RBZ,
\begin{eqnarray}
g(k,q,\omega)= \sum_{i=1}^4 [G_{ii}(k,k',
s_i\omega)-G^*_{ii}(k',k,s_i\omega)]
\end{eqnarray}
where $s_i=1$ for $i=1,2$ and $s_i=-1$ for $i=3,4$. If $k'$ is
not in the RBZ, let $k''=k+q-Q$. For this case
\begin{eqnarray}
 g(k,q,\omega) &=&\sum_{i=1,3}
[G_{i,i+1}(k,k'',s_i\omega)-G^*_{i,i+1}(k'',k,s_i\omega) \nonumber
\\& +&G_{i+1,i}(k,k'',s_i\omega)-G^*_{i+1,i}(k'',k,s_i\omega)]
\label{g3}
\end{eqnarray}
Here $G(k_1,k_2,\omega)$ is obtained by analytical continuation $i
\omega_n \rightarrow \omega+i \delta$ of $G(k_1,k_2,i \omega_n)$
from imaginary frequencies to real frequencies. The complexity of
the above formula stems from translational symmetry breaking
in the presence of DDW order.

\section{Numerical calculation and discussion}
We compute the local density of states $\rho(q,\omega)$ using
the $4\times 4$ impurity scattering matrices. For
potential scattering given by a $\delta$-function, they are
\begin{eqnarray}
V(k,k')=V_N\left( \begin{array}{cccc}
1 &1 &0& 0\\
1 & 1 &0& 0\\
0 &0& -1&-1\\
0 &0 &-1& -1
\end{array} \right ).
\end{eqnarray}
for a non-magnetic impurity, and
 \begin{eqnarray}
V(k,k')=V_M\left( \begin{array}{cccc}
1 &1 &0& 0\\
1 & 1 &0& 0\\
0 &0& 1&1\\
0 &0 &1& 1
\end{array} \right ).
\end{eqnarray}
for a magnetic impurity.

The   results  are reported for the representative values:
$V_N=V_M=0.1 \; \text{eV}$. For the  band dispersion, we choose $t=-1.2
\; \text{eV}$ and  $t'=0.36\; \text{eV}$. The chemical
potential is selected to be equal to $-0.36 \; \text{eV}$. The imaginary
part of the energy $\delta=0.5 \;  \text{meV}$ is used for the entire
numerical calculation. We have checked that the results are
unchanged  for smaller values of $\delta$. Following
Ref.~\onlinecite{Wang03},   a $400 \times 400$ lattice is used in our
analysis, and the results are displayed in the $(-\pi,\pi) \times
(-\pi,\pi)$ interval on a $49 \times 49$ grid for any given
frequency. The choice of these
parameters is representative. We have repeated
our calculations for a number of different
set of parameters; the conclusions remain unchanged.

\subsection{The DSC state}  First we analyze the interference
patterns in a pure
superconducting state with $\Delta_0=25 {\rm meV}$. This has also
been studied in Ref.~\onlinecite{Wang03} for a different bare band
structure.  The constant energy contour plots are shown in the
Fig.~\ref{surface}(b) where we label the wavevectors, which are
expected to be associated with the peaks in the interference
patterns. In Fig.~\ref{result1}, we show a comparison between our
results and the results in Ref.~\onlinecite{Wang03}, where a different
band structure \cite{Norman94} which is flatter near $(\pi,0)$ is used.
Some of the features in our results are the same as in
Ref.~\onlinecite{Wang03}, but there are also several clear
differences.

For the case of a
nonmagnetic impurity, the peaks associated with the wavevectors
numbered 3, 4, and 7  are observed for a large energy range.
However, the intensity is high  mainly along the diagonal or
symmetrically about the diagonal, contrary to the results in
Ref.~\onlinecite{Wang03}, where regions of high intensity emerge along the
$(0,\pm 1)$ and $(\pm1,0)$ directions when the energy $|\omega|$
reaches  $15 \text{meV}$. The peak associated with the wave vector
numbered 1 appears for some energies, but with weaker intensity
than the peaks associated with the above three wave vectors.

For the case of a magnetic impurity, there are fewer features,
so it is more difficult to relate the observed intensity to
the scattering from the tips of the banana-shaped contours.

\subsection{The mixed DSC and DDW state}

If DDW order is the origin of the pseudogap phase, the mixed state
should describe the  underdoped cuprates at low temperatures. In
Fig.~\ref{surface}(d) we sketch  the equal energy contours for the
band structure of a mixed DDW-DSC state. For energies lower than
the DSC gap the equal energy contours are now four pairs of
bananas due to the doubling of unit cell by the DDW order,  and
the regions of high DOS are situated at their tips. However, for
energies larger than the DSC gap,  the equal energy contours
become elliptical, which is characteristic to the DDW
state.

In Fig.~\ref{mix1}, we plot our results for the mixed state for
nonmagnetic impurity scattering. We note that for energies lower
than the DSC gap, we can identify exact features characteristic to
the pure DSC state. The patterns are almost identical to those
obtained in a pure DSC state with the same DSC gap $\Delta_0$.
In particular
the peaks along the $(\pi,\pi)$ axis disperse with energy in the
expected fashion. However, they reach the corners of the BZ at
energy comparable to the DSC order parameter $\Delta_0$,
and not to the full gap of the system $\Delta_{t}
=\sqrt{W_0^2+\Delta_0^2}$. Above the
energy $\Delta_0$, a whole range of different features emerge,
which are characteristic to the DDW state.
For comparison, in Fig.~\ref{mix2} we plot the corresponding
results for the case of a pure DSC state with a gap equal to the
total gap of the mixed state $\Delta_t$. Indeed, the peaks in
the pure DSC state disperse much slower with energy and
can be observed at higher energies than in the mixed state.
Based on these
observations we expect the FT-STS measurements to be an
experimental tool to observe the coexistence of DDW order with
DSC order in the underdoped cuprates. If this state is
indeed a coexisting DDW and DSC state, measurements done at
various energies should reveal that the diagonal peaks in the
spectra should approach the corners of the BZ well before
the full gap is reached. Also, entirely new features should arise
in the STM spectra for energies above the energy for which the
peaks have reached the corners. The new features would be similar
to what one would expect to see in a pure DDW state. Broad regions
of high intensity are expected rather than sharp peaks,
as we will discuss in the next subsection. When the doping is
increased, since the total gap $\Delta_t$ is roughly constant for
different doping levels, the DDW gap decreases and the DSC gap
increases. The energy at which the transition described above
happens in the interference patterns is thus expected to get
closer and closer to the total gap and eventually equal to it
when the DDW gap vanishes and the DDW ordering disappears.

\subsection{The DDW state}

We now turn to a pure DDW state. In Fig. ~\ref{surface}(c) we
plot a typical band structure of a pure DDW state.  We note that
the equal energy contours are now elliptical and we therefore
expect the high DOS regions to be situated at their tips, though,
due to the smaller curvature of these contours, the distinction
between the high and low regions of DOS are not as strong as
in the case of DSC. In Fig.~\ref{ddw}, we plot the corresponding
LDOS results for $W_0=25\;\text{meV}$, and $W_0=40\;\text{meV}$.

Even though one can still track the presence of peaks, they are
not as pronounced. This may
be the result of a more uniform DOS. The position of some of the
peaks (such as some
along the ($\pi, \pi$) direction and symmetrically about it, and
some along the ($\pi,0$) direction) may be traced back to the
positions of the tips of the elliptical contours in the band
structure. These peaks are present in most of the pictures,
but their intensities vary.

Also, for the chosen band structure, peaks located around $(\pm
\frac{\pi}{4}, 0)$ and $(0,\pm \frac{\pi}{4})$ can  be found over a
broad range of energy. They are present in both the $W_0=40\:\text{meV}$ and
$W_0=25\:\text{meV}$ data. In particular, their positions hardly change
with energy for positive energies. We note that
their peculiar positions of roughly $(\pm \frac{\pi}{4}, 0)$
and $(0,\pm \frac{\pi}{4})$ is  an effect of the
particular set of parameters in the band structure.
The origin of these peaks is the
scattering indicated by the arrow in
Fig.~\ref{surface}(c). Since the equal energy contours
may change drastically with a change of parameters, so may the
position of the peaks.
We also note that, since the equal energy  contours
change very little with energy for positive energies,
the peaks also do not move
when the energy is varied.

Many other peaks however cannot be
explained by simple band structure arguments, and their positions change
drastically from one plot to the next.

The predominant feature for the DDW state is thus the presence of broadly
distributed scattering points in the interference patterns at  low energy.
However, lines with relatively high intensity are also present in our data.
Such lines actually occur in all of the states considered.
Furthermore, the intensity and size of these lines
vary strongly with even slight changes of parameters.
This is in contrast to the statement in Ref.~\onlinecite{Pereg-Barnea03}
that the presence of high intensity lines in the spectrum is a
characteristic of the DDW state.
However, since the equal energy contours are elliptical, the scattering
patterns in the DDW states are much more likely to be affected by
coherence factors than the scattering patterns
in other states. Because of this effect,  the
comparison between DDW and DSC in  Ref.~\onlinecite{Pereg-Barnea03},
based on  Born approximation,  should not be considered as generic,
although we generally agree that the interference patterns are
different in these states. We will discuss this in more detail
in the next subsection.

As a final observation, we note that if we tune the parameters
such that electron pockets around $(\pm \pi,0)$ and $(0,\pm \pi)$
appear in the DDW  band structure, our calculation shows circles
of high intensity in the BZ corresponding to electrons with
different momenta scattering off each other. This appears to be
the dominant feature. However, the presence of such electron
pockets is not consistent with ARPES data\cite{ARPES} and we
do not include the results for this situation in the present work.

\subsection{ Discussion of the coherence factors  }

The interference patterns not only depend
on the band structure and on the order parameters, but
also on the coherence factors. This dependence on coherence factors
has quite complex effects and
makes the interference patterns sensitive to
many details.

First, the coherence factors depend on the strength and
the type of the impurity scattering.  In Ref.~\onlinecite{Wang03}, the
interference patterns from nonmagnetic and magnetic
impurities were shown to be so different that entirely different peaks were
observed, even for pure $s$-wave scattering.
Clearly, similar or more pronounced effects may arise
in the case of non $s$-wave scattering.
Furthermore, the coherence factors depend on the impurity scattering strength.
This may be implied from the sensitivity of the interference patterns
to whether we use the Born approximation or a full
T-matrix calculation for a reasonable impurity strength.

Second, the coherence factors are also strongly sensitive to the
band structure and order parameters.
The band structure dependence
causes the dramatic difference
between our results for the DSC state
and the results in Ref.~\onlinecite{Wang03}.
The flatness of the band structure of Ref.~\onlinecite{Norman94}
leads to differences in the DOS and also to important
differences in the coherence factors.
In the DDW state, the effect of coherence factors
is more dramatic than in the DSC states, as seen from the
plot of our results for identical parameter sets save for
the DDW gaps. Though the band structures for the two chosen
DDW gaps are very similar, as seen in
Fig.~\ref{ddw} there are considerable differences between the
corresponding FT-STS patterns.

Third, the coherence factors are sensitive to  energy. This has
been shown consistently in all our calculations. Even for
similar band structures, some peaks can
only be identified for some ranges of energy, while others appear or
disappear in a manner counterintuitive
to expectations based on band structure.

Finally, the coherence factors should also be expected to
depend on the correlation and distribution of impurities in
the materials.

The sensitivity of the  coherence factors to so many details makes it
difficult to extract information about the dispersion of
quasiparticles from the interference patterns. It is rather hard
to identify a feature that can be traced in a broad energy range.
This is particularly true for the DDW state, where
the quasiparticle scattering interference is
much more sensitive to the changes of the parameters in the model.

\section{Conclusions}
In this paper we have analyzed the effects of scattering from a
single impurity on FT-STS in a high $T_c$ superconductor with DDW
ordering. For the case of a pure superconducting state, we
recovered results similar to Ref.~\onlinecite{Wang03}. We note the
presence of peaks dispersing with energy in the expected fashion.
However, some important features are strongly dependent on the
band structure and on other parameters in the model even if the
banana-shaped equal-energy plots are similar.   In the case of a
mixed DDW-DSC state, we note that the DSC order dominates the
interference patterns at energies lower than the DSC gap
$\Delta_0$, while for higher energies,
 a transition to a spectra characteristic of the DDW phase occurs.
This is marked by a sudden change of the general features of the
STM spectra around an energy $E=\Delta_0$. In particular, we show that the
dispersion of the diagonal peaks in the spectrum can be used to
identify this transition. In the underdoped regime,
the DSC gap $\Delta_0$ is reduced when the doping decreases,
so the energy at which the transition
happens  should also be reduced.
Based on this observation, we predict that
the STM measurements in the underdoped cuprates below $T_c$ may
reveal important information about the nature of the gap.
In the pure DDW state the results are more complex; the
peaks are more broadly spaced and not so pronounced and are more
sensitive to changes of parameters.  It would be interesting to
see if STM experiments in the pseudogap regime can provide new
insight into the problem.

\section{Acknowledgments}
C. B. has been supported by the NSF
under Grant No. DMR-9985255, and also by funds from the A. P.
Sloan Foundation and the Packard Foundation. S. C. has been supported by the NSF
under Grant No. DMR-9971138.
JP was supported by the funds from
the David Saxon chair at UCLA.
C. N. has been supported by the NSF under Grant No. DMR-9983544
and the A.P. Sloan Foundation.

\begin{figure*}
\begin{center}
\includegraphics[width=7in]{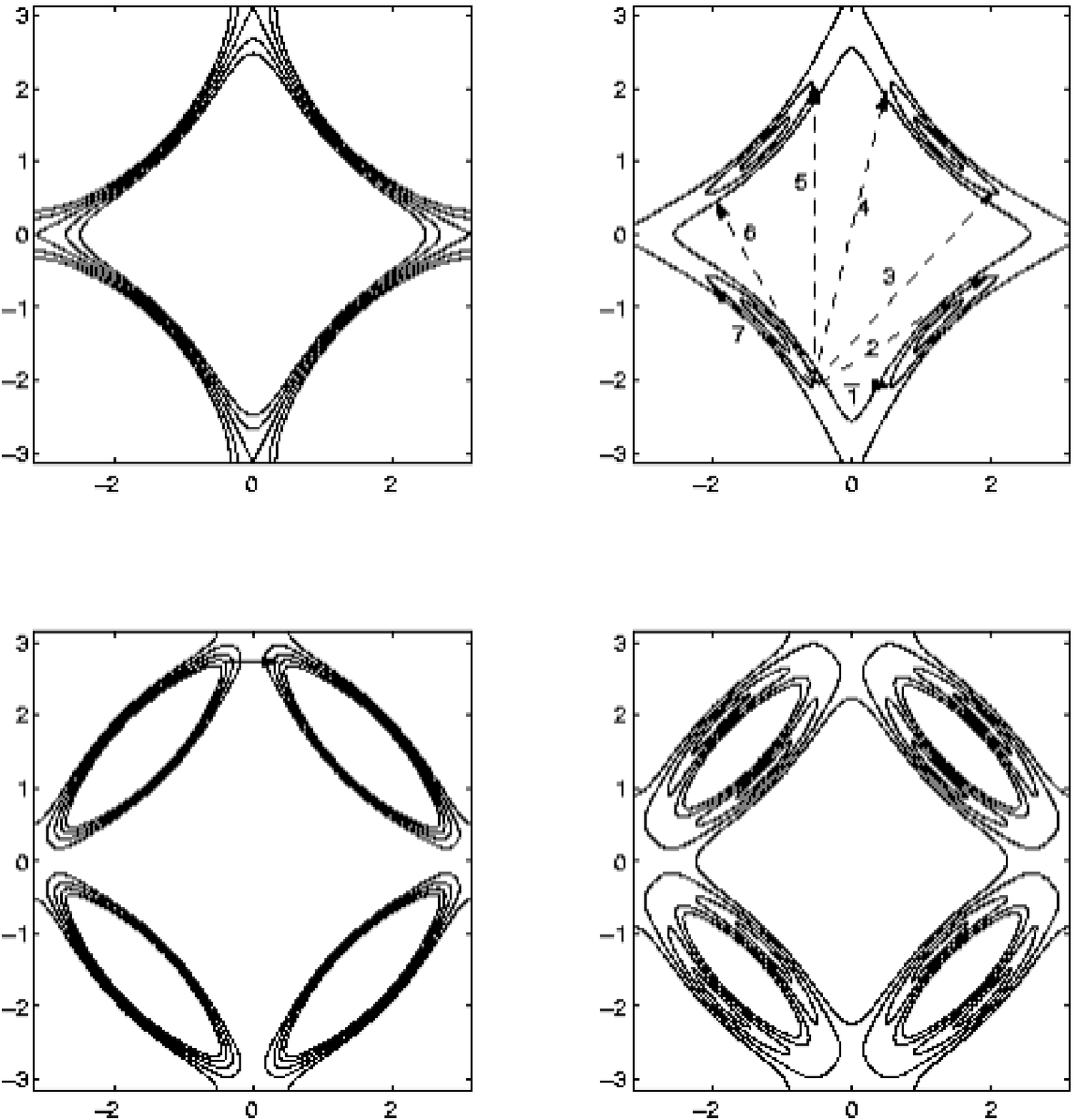}
\end{center}
\caption{Equal energy contour plots in
different states: (a) normal state without any order, (b)
$d$-wave superconducting state (DSC), (c) $d$-density wave state (DDW), and
(d) Mixed DSC and DDW state (DSC-DDW). }
\label{surface}
\end{figure*}

\begin{figure*}
\begin{center}
\includegraphics[width=7in]{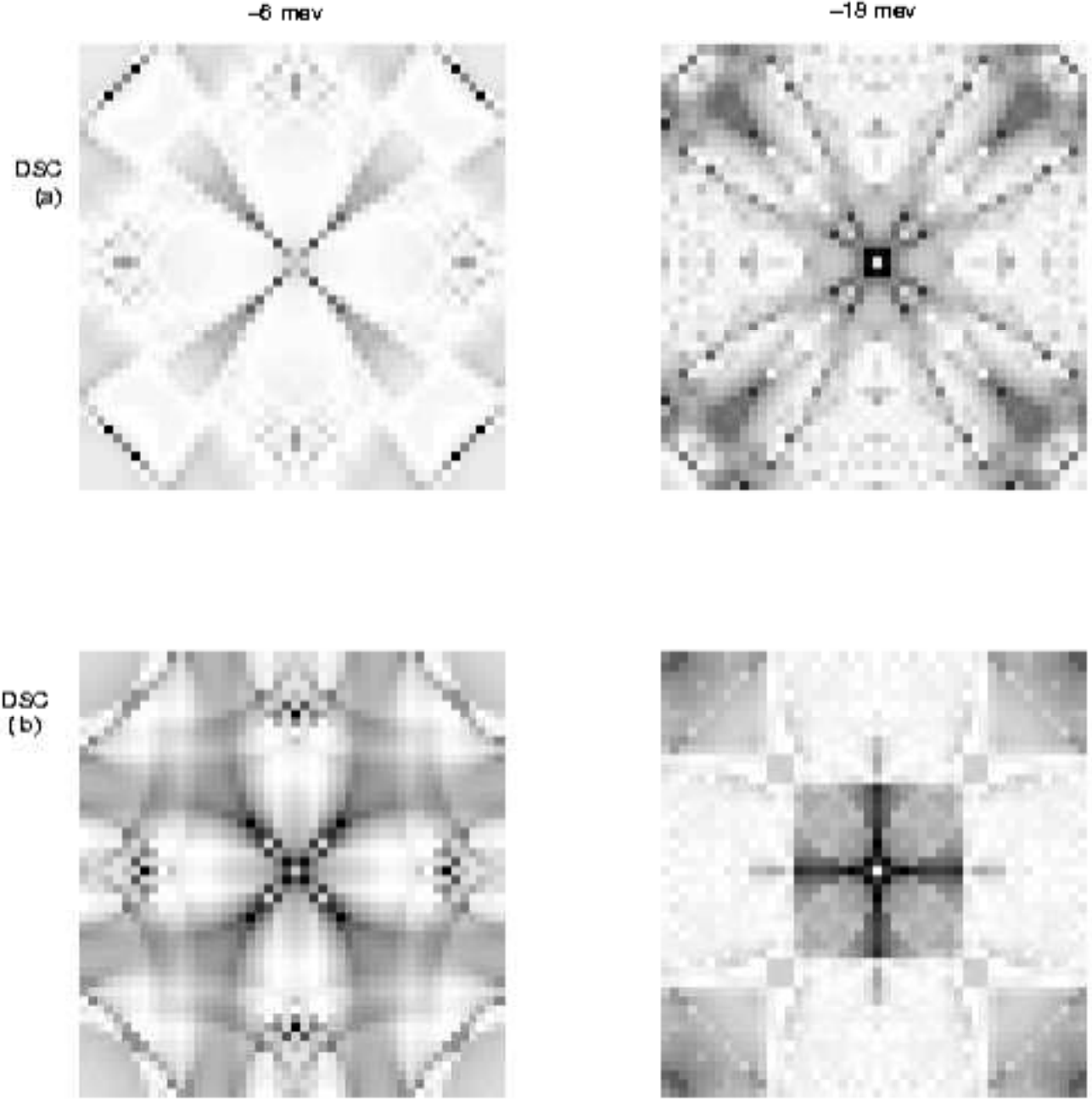}
\end{center}
\caption{Fourier transform of STM spectra: (a) in a pure DSC state with the
$t-t'$ band structure and  $\Delta_0=25$ \text{meV}, (b) in a pure DSC
state with the band structure in Ref.~ \onlinecite{Norman94} and $\Delta_0=25$
\text{meV} and at energies equal
to $-6 \text{meV}$ and $-18 \text{meV}$, for nonmagnetic impurity scattering
($V_N=100 \text{meV}$). The differences between (a) and (b) reflect
the sensitivity of FT-STS spectra to band structure details.}
\label{result1}
\end{figure*}

\begin{figure*}
\begin{center}
\includegraphics[width=7in]{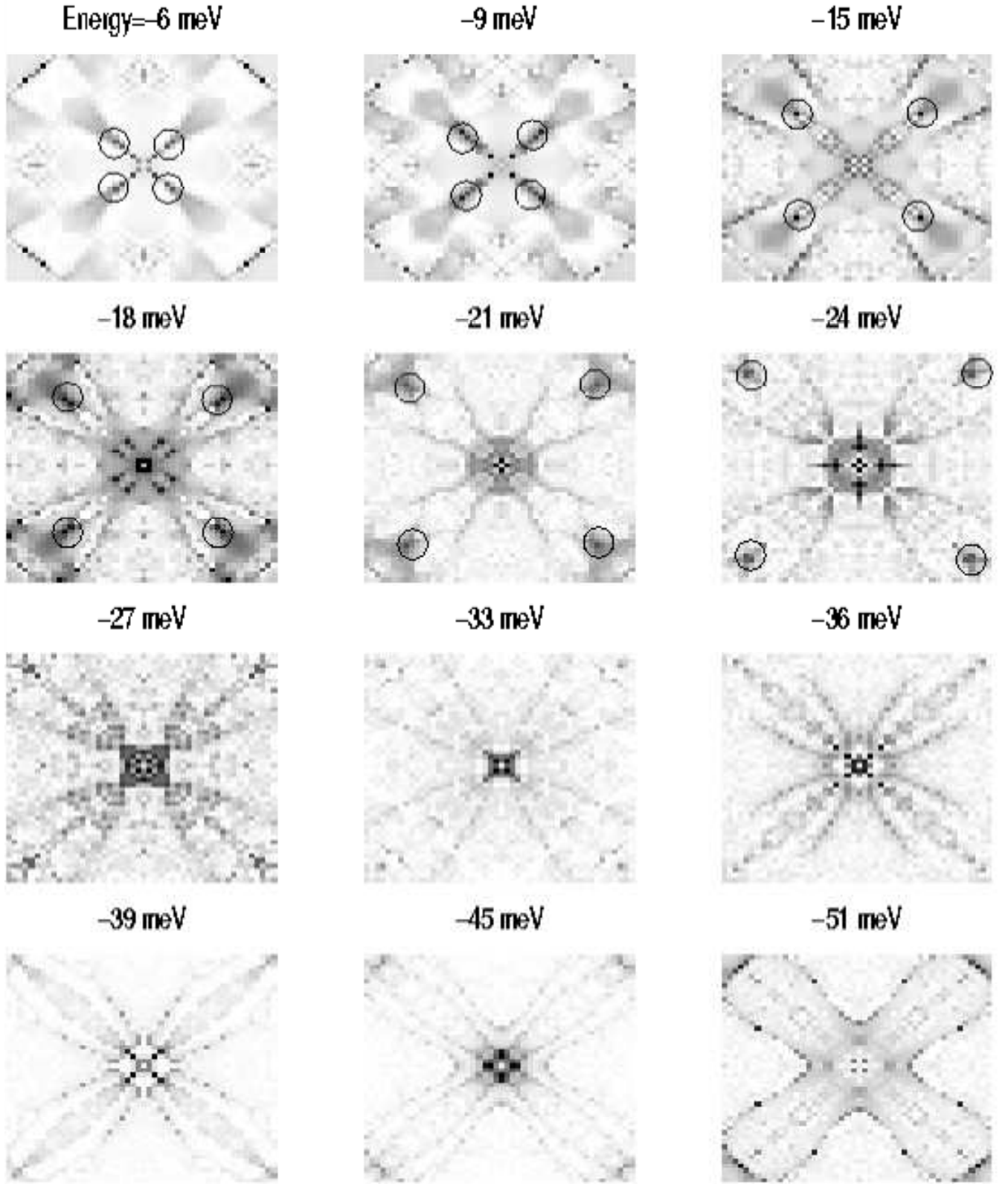}
\end{center}
\caption{Fourier transform of STM
in the coexisting DSC-DDW state with the $t-t'$ band
structure, $\Delta_0=25 \text{meV}$ and $W_0=40 \text{meV}$,
for nonmagnetic impurity scattering
($V_N=100 \text{meV}$). The peaks along the directions $(\pm\pi,\pm\pi)$
are marked by circles. As noted, these peaks
get close to the corners of the BZ at energies equal to the
DSC gap $\Delta_0$, and not to the total gap, which is $\Delta_t \approx 47 \text{meV}$.
Also, the general features of
the spectra change dramatically at
energies larger than  $\Delta_0$. }
\label{mix1}
\end{figure*}

\begin{figure*}
\begin{center}
\includegraphics[width=7in]{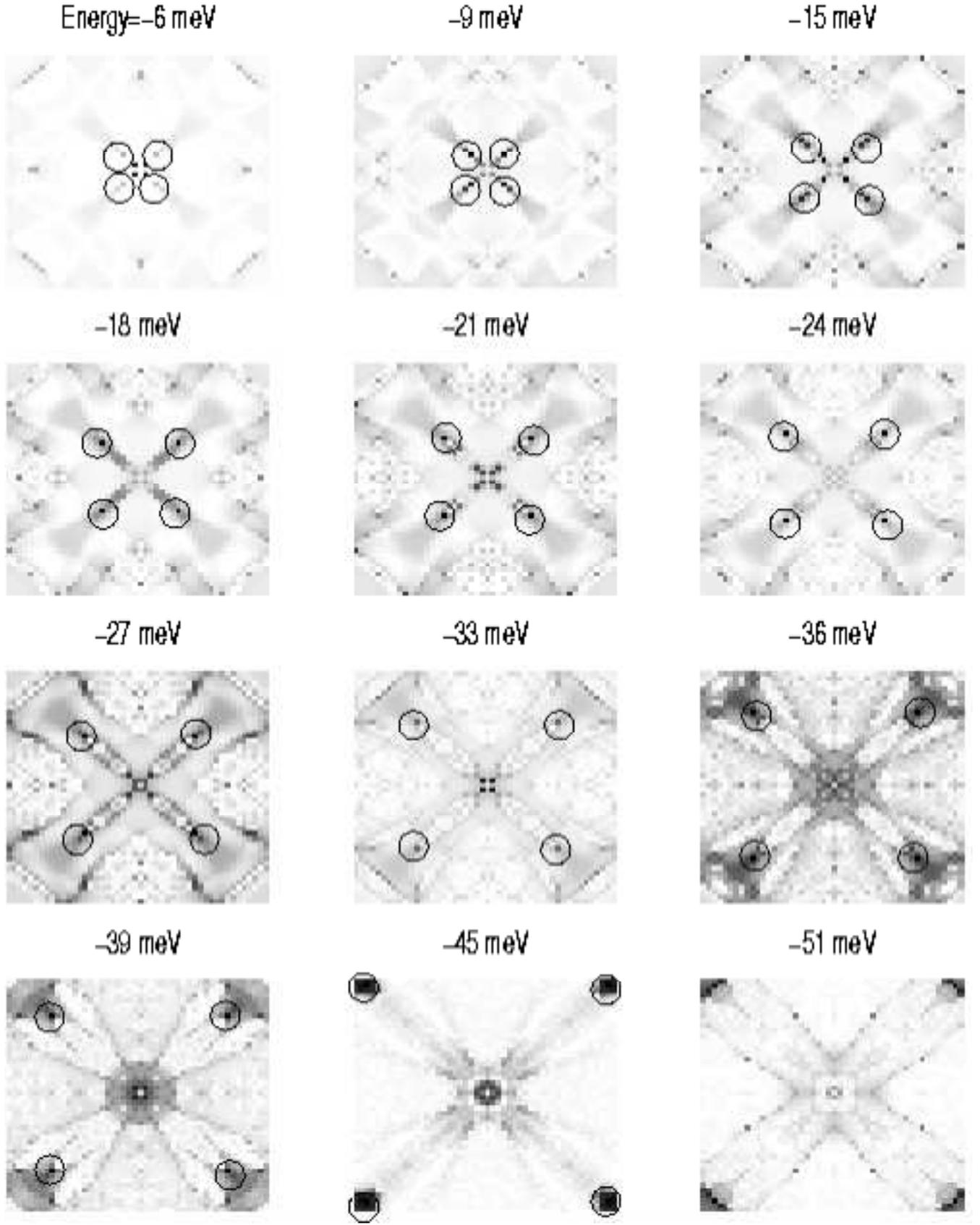}
\end{center}
\caption{Fourier transform STM spectra
in a pure DSC state with the $t-t'$ band
structure and $\Delta_0=47 \text{meV}$,
for nonmagnetic impurity scattering
$V_N=100 \text{meV}$. Again, the peaks along the directions $(\pm\pi,\pm\pi)$
are marked by circles. In this case, the peaks get close
to the corners of the BZ for energies equal to the DSC gap
$\Delta_0=47 \text{meV}$.}
\label{mix2}
\end{figure*}

\begin{figure*}
\begin{center}
\includegraphics[width=7in]{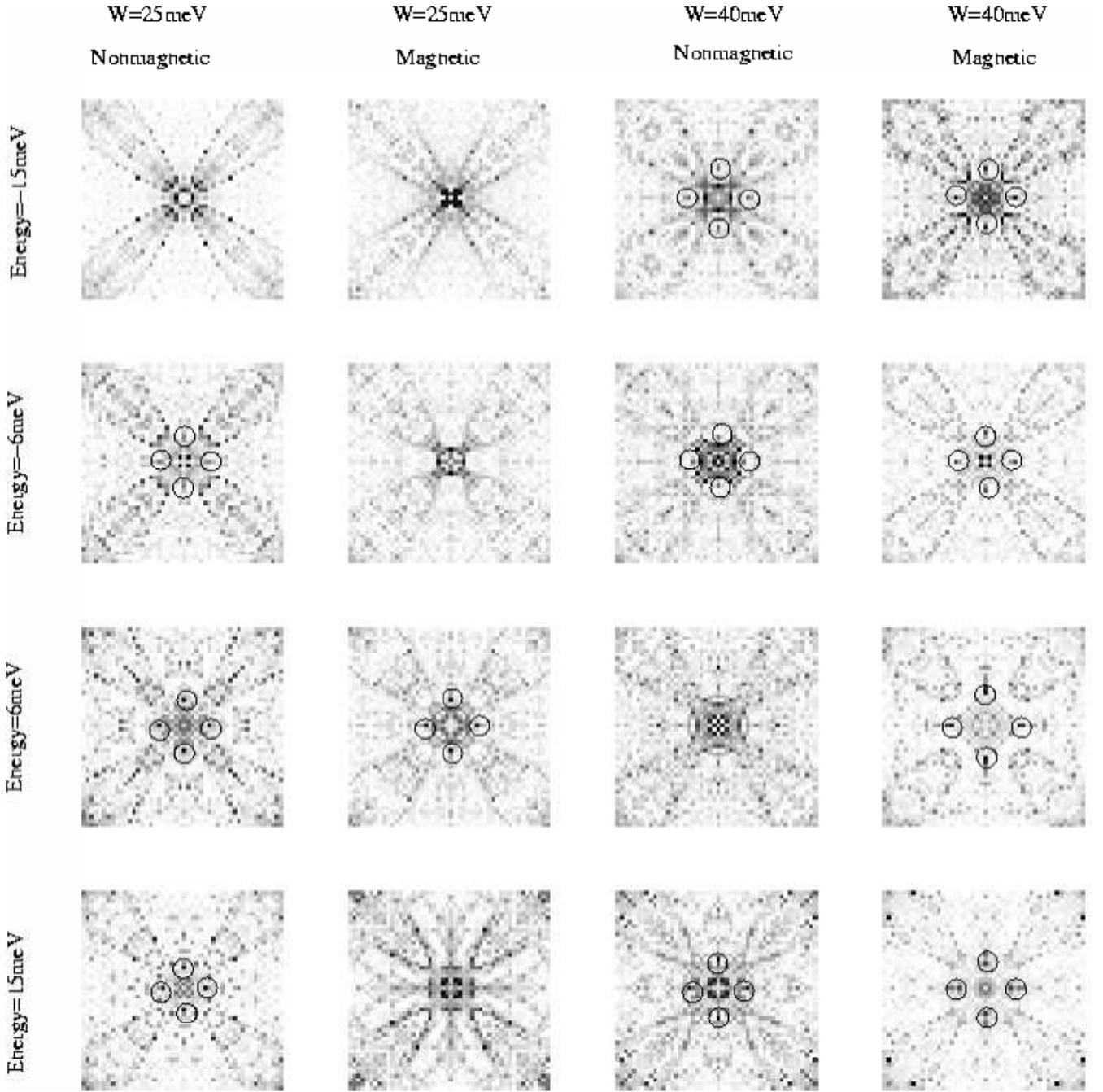}
\end{center}
\caption{Fourier transform STM spectra in the pure DDW state with
the $ t-t'$ band structure  and $W_0=40 \text{meV}$, for
non-magnetic ($V_N=100 \text{meV}$) and magnetic ($V_M=100
\text{meV}$) impurity scattering. The peaks at $(0,\pm \pi/4)$ and
$(\pm \pi/4,0)$ are marked by circles and can be observed for a
large range of energies.} \label{ddw}
\end{figure*}

\end{document}